\documentclass{ws-procs975x65}
\usepackage{amsmath}
\usepackage{amsfonts}
\usepackage{amssymb}
\usepackage[dvips]{epsfig}

\begin{document}

\title{Consequences of observational uncertainties on the detection of cosmic topology}

\author{B. Mota,\ \  M.J. Rebou\c{c}as}

\address{Centro Brasileiro de Pesquisas F\'{\i}sicas\\
Rua Dr.\ Xavier Sigaud, 150, \\ CEP 22290-180 Rio de Janeiro --
RJ, Brazil\\ E-mail: brunom@cbpf.br, reboucas@cbpf.br}

\author{R. Tavakol}

\address{Astronomy Unit, School of Mathematical Sciences, \\
Queen Mary, University of London, \\ Mile End Road, London E1 4NS,
UK\\ E-mail: r.tavakol@qmul.ac.uk}

\maketitle

It is well known that as a consequence of local nature of general
relativity, the global topology of space-time remains undetermined
by the Einstein's field equations. This coupled with the enormous
recent increase in high resolution cosmological observations has
led to a great deal of interest in the possibility that the
universe may possess compact spatial sections with a non-trivial
topology (see for example Refs. \refcite{CosmicTop} and
\refcite{Revs}). These observations have also shown that the
spatial curvature is very small (and possibly $0$).\cite{WMAP}
Whatever the nature of cosmic topology may turn out to be, the
issue of its detectability is of fundamental importance.

Motivated by these observational results, a study was recently
made of the question of detectability of the cosmic topology in
nearly flat universes. It was demonstrated that as
$\Omega_{0}\rightarrow1$ increasing families of possible
topologies become undetectable by methods based on image (or
pattern) repetitions (see Refs.
\refcite{grt2001a}~--~\refcite{NEWweeks2}). However, measurements
of the density parameters unavoidably involve observational
uncertainties, and therefore any study of the detectability of the
cosmic topology should take such uncertainties into account.

In a recent paper,\cite{our} we studied the sensitivity of the
detectability of cosmic topology to the uncertainties in the
density parameters, using two complementary methods. Here we
briefly summarise some of those results.
%These methods
%provide respectively sufficient and necessary conditions for
%undetectability of cosmic topology. From one of these methods (the
%secant line method) it is possible to derive an exact closed form
%for a set of sufficient conditions for undetectability of cosmic
%topology of nearly flat universes. It can be further shown
%(numerically) that the converses of these conditions are (to a
%good approximation) also sufficient to determine detectability in
%principle.

%Here we give a brief account of this method and summarise some
%examples of its use to determine undetectability of cosmic
%topology of classes of manifolds.
%
As in standard cosmology we assume the universe is modelled by a
$4$-manifold $\mathcal{M}=\mathcal{R}\times M$, with a locally
isotropic and homogeneous Robertson-Walker (RW) metric, with a
matter-energy content well approximated by dust (of density
$\rho_{m}$) plus a cosmological constant $\Lambda$, with
associated fractional densities $\Omega_{m}=8\pi G\rho
_{m}\,/\,(3H^{2})$ and
$\Omega_{\Lambda}\equiv\Lambda\,c^{2}/\,(3H^{2})$, and
$\Omega_{0}=\Omega_{m}+\Omega_{\Lambda}$. We also assume a small
but non-zero curvature, since a flat universe has no preferred
length scale, and therefore the cosmological parameters impose no
constraints on observable candidate manifolds. The
redshift-(comoving)-distance relation in units of curvature radius
then takes the form
\begin{equation}
\chi(z)=\sqrt{|1-\Omega_{0}|}\int_{0}^{z}\left[  (1+x)^{3}\Omega_{m0}%
+\Omega_{\Lambda0}-(1+x)^{2}(\Omega_{0}-1)\right]  ^{-\frac{1}{2}%
}dx\;,\label{chi(z)}%
\end{equation}

%In order to study the (possibly non-trivial) topology of the
To study the topology of the spatial sections $M$ of the universe,
we need a topological invariant length that could be put into
correspondence with depth of surveys. We employ the injectivity
radius $r_{inj}$, the radius of the smallest sphere 'inscribable'
in $M$, which is defined as half the length $\ell_{M}$ of the
smallest closed geodesics, $r_{inj}=\frac{\ell_{M}}{2}$ (for
details see~\cite{grt2001a}). A manifold $M$ is then detectable in
principle in a survey of depth $z$ if the density parameters are
such that $\chi(z)> r_{inj}$. If $\chi(z)\leq r_{inj}$ then the
topology is undetectable by any pattern repetition method.

This question can be restated in terms of countour lines in the
$\Omega _{m0}-\Omega_{\Lambda0}$ parametric plane. For any given
manifold $M$ with injectivity radius $r_{inj}^{~M}$ and fixed
survey depth $z_{obs}$, we can define the contour curve
$\chi(z_{obs},\Omega_{m0},\Omega_{\Lambda0})=$ $r_{inj}^{~M}$ .
This curve lies in either of two regions: the positive curvature
($\Omega_{0}>1$) or the negative curvature ($\Omega_{0}<1$)
semi-planes, depending on whether the manifold is respectively
spherical or hyperbolic in nature. The contour curve further
subdivides its semi-plane in a region where the topology is
undetectable ($\chi_{obs}<r_{inj}^{~M}$), and a region where the
topology is detectable in principle ($\chi_{obs}\geq
r_{inj}^{~M}$). Therefore, given this curve, it would be possible
to determine the (un)detectability of any given nonflat manifold
for a range of density parameters. The question then becomes how
to determine these countour curves.

In a recent work\cite{our}, we developed two complementary linear
approximations for countour curves, one that necessarily
overestimates detectability (by approximating the countour curve
by its tangent line, and therefore named the tangent line method),
and one that underestimates detectability (the secant line method,
which approximates the countour curve by the line connecting its
the extremes,$(\widetilde{\Omega}_{m0},0)$ and
$(0,\widetilde{\Omega}_{\Lambda0})$ ). We shall not describe
either method in detail here, but see Figure 1 for a qualitative
description.

We found the numerical results from both methods to be in good
agreement, for density values compatible with current
observations,\cite{WMAP} thus demonstrating that they provide good
approximations to the contour curve. Furthermore, the equation for
the secant line can be obtained analytically in the limit
$z\rightarrow \infty$. With this equation one can write the
following inequalities, which is useful to state the conditions
for (un)detectability of cosmic topology:

\parbox{10cm}{\begin{eqnarray*}
\cosh^{2} \,(\, \frac{r_{inj}^{\:M}}{2}\,)\; \Omega_{m0} +
\Omega_{\Lambda0} & >& 1\;, \quad \mathrm{for} \;\quad
\Omega_{0}<1\;,  \nonumber\\ \cos^{2}\,(\,
\frac{r_{inj}^{\:M}}{2}\,)\; \Omega_{m0}+ \Omega_{\Lambda0} & <
&1\;, \quad \mathrm{for} \;\quad \Omega_{0}>1\;.
                \end{eqnarray*}}  \hfill
\parbox{1cm}{\begin{eqnarray}   \label{analit}  \end{eqnarray}}

This last result is of particular interest, because it allows the
study of the detectability of topology not only in individual
manifolds, but also in whole classes of manifolds.\cite{our}
\begin{figure}[tbh]
\centerline{\def\epsfsize#1#2{0.55#1}\epsffile{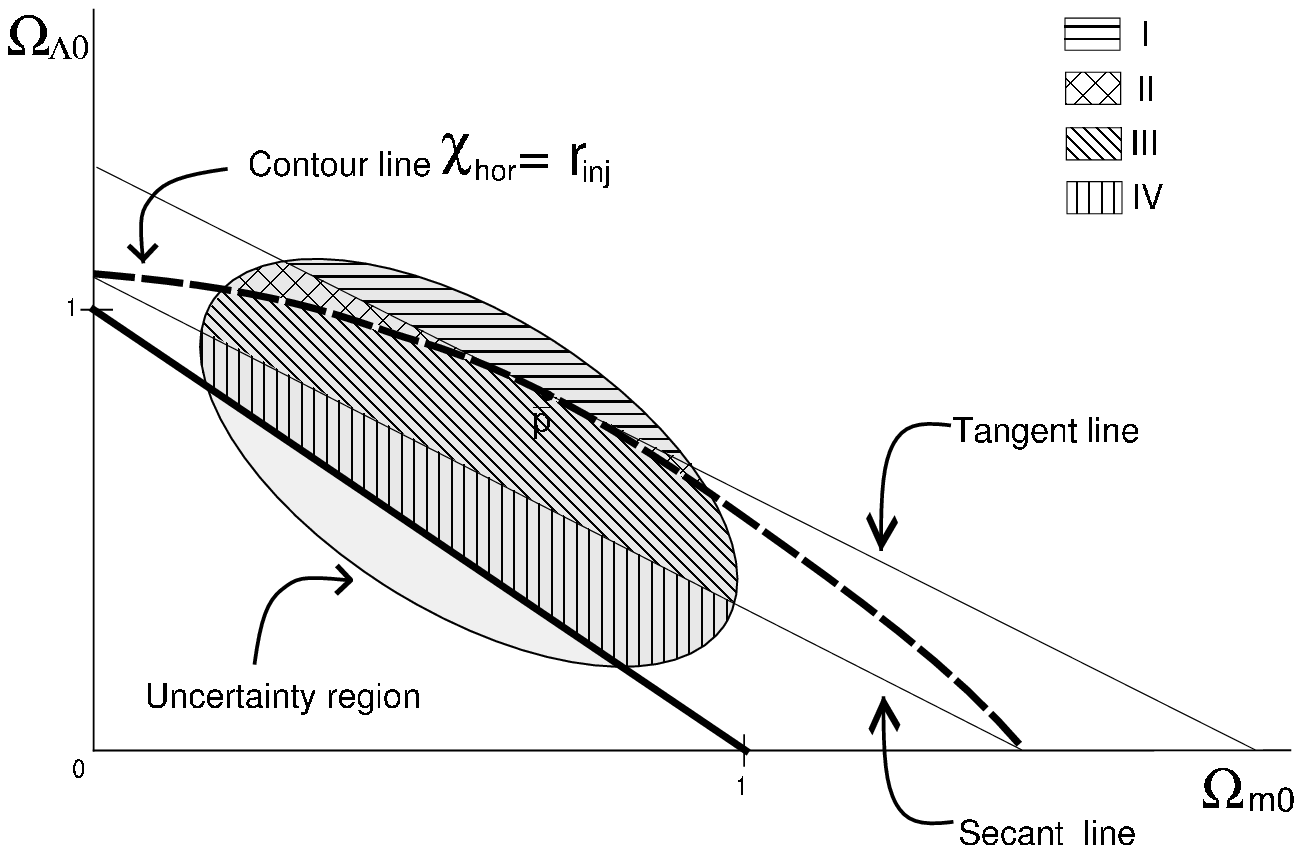}}\caption{A
schematic representation of the secant line (SL) and tangent line
(TL) methods. The convexity of the contour curve for
$\Omega_{0}>1$ can be proven analytically. The topology is shown
to be detectable in principle in region I by the TL method, and
undetectable in region IV by the SL method. Regions II and III are
respectively detectable and undetectable, but are not
discriminated by either methods. But II and III are only a very
small fraction of the uncertainty region $\,\mathcal{U}\,$ for
manifolds whose contour curves intersect the uncertainty region.}%
\label{SchemFig}%
\end{figure}

A consequence of these results is that the closed form
inequalities~(\ref{analit}) can be seen, to a very good
approximation, as establishing conditions for detectability in
principle as well, as can be shown by comparison with numerical
values obtained from both methods for $z=1100$. For high redshifts
we can therefore use~(\ref{analit}) to separate the parameter
plane into undetectable and detectable sub-regions with great
accuracy. The closed form of the inequalities makes its
application quite straightforward and potentially more useful.

Finally even though we have used a $\Lambda$CDM framework, similar
methods could be developed for other cosmological
models~\cite{MaklerMG10} with different redshift-distance
relations in order to obtain conditions for undetectabilty of
cosmic topology.\newline

%%%%%%%%%%%%%%%%%%%%%%%%%%%%%%%

We thank CNPq and FAPERJ for financial support.

\end{document}